\documentclass[twocolumn,aps,prl,superscriptaddress,showpacs,amsfonts,amsmath,amssymb]{revtex4}
\usepackage[dvips]{graphicx}
\usepackage{epsfig}
\newcommand{\rhot}{\tilde{\rho}}
\newcommand{\tw}{t_{\rm w}}
\newcommand{\be}{\begin{equation}}
\newcommand{\ee}{\end{equation}}

\begin{document}

\title{Spatially heterogeneous dynamics in granular compaction}

\author{Alexandre Lef\`evre}

\affiliation{Rudolf Peierls Centre for Theoretical Physics, University
of Oxford, 1 Keble Road, Oxford, OX1 3NP, UK}

\author{Ludovic Berthier}

\affiliation{Laboratoire des Verres UMR 5587, Universit\'e Montpellier
II and CNRS, 34095 Montpellier, France}

\author{Robin Stinchcombe}

\affiliation{Rudolf Peierls Centre for Theoretical Physics, University
of Oxford, 1 Keble Road, Oxford, OX1 3NP, UK}

\date{\today}

\begin{abstract}
We prove the emergence of spatially correlated dynamics 
in slowly compacting dense granular media by analyzing analytically  
and numerically multi-point correlation functions in a simple 
particle model characterized by slow non-equilibrium dynamics.
We show that the logarithmically slow 
dynamics at large times is accompanied by 
spatially extended dynamic structures that resemble 
the ones observed in glass-forming liquids and 
dense colloidal suspensions. 
This suggests that dynamic heterogeneity is another key common
feature present in very different jamming materials.
\end{abstract}


\maketitle

When a pile of grains is gently shaken, its volume fraction 
increases so slowly that the process hardly becomes 
stationary on experimental timescales~\cite{kni,bideau}. 
This is reminiscent of the slow relaxation observed
in glass-formers, as noted long ago~\cite{struik}.
From a more fundamental point of view, 
it is tempting to build upon analogies 
and suggest that granular media, glasses, 
and other jamming systems can be described by common 
theoretical approaches~\cite{jamming}. 
In recent years, several aspects of glasses and granular 
media have been studied with similar approaches.
Static structures have been studied to understand the relevance
of the network of force chains between grains or atoms
to dynamics of the jammed state~\cite{force}. 
Also, since both grains and glasses undergo 
non-equilibrium `glassy' dynamics, the idea that an effective thermodynamics
can be used has received considerable 
attention~\cite{reviewaging,teff}.

In this work we also transfer knowledge from
one field to the other and show that the glassy dynamics 
of granular media is characterized by the appearance
of spatio-temporal structures similar 
to the ones described as dynamic heterogeneity in 
glass-formers~\cite{dh} and dense colloidal  suspensions~\cite{wee}.
Dynamic heterogeneity 
is believed to play a crucial role in the glass formation, 
and forms the core of recent theoretical descriptions~\cite{kcm}.
Physically, it stems from the existence of spatial correlations 
in the local dynamics that extend 
beyond the ones revealed by static pair correlations.
To study dynamic heterogeneity,
correlators that probe more than two points in space and time
have to be considered~\cite{dh,kcm,mayer}. 
These spatial fluctuations
have never been studied in models or experiments 
on granular compaction, although caged particle dynamics
was recently studied in a sheared system~\cite{mar}.    
Here we prove the emergence of
large dynamic lengthscales in a particle model which is a variant 
of the parking lot model, introduced 
and studied in detail in the context of 
granular compaction~\cite{kra,sti,dep,kol,tal}. 
We take advantage of its relative simplicity to compute
analytically multi-point correlations
studied in glass-formers and confirm our results
by numerical simulations. 
   
We consider a variant of the parking lot model~\cite{kra}
introduced in Ref.~\cite{sti}. 
The model is a one-dimensional process  
in which hard blocks of unit size are first irreversibly deposited 
at random positions on a line of linear size $L$
until no place is available for more depositions. 
Timescales are counted from the end of the deposition 
process which corresponds to $\tw=0$. 
In a second step, particles are allowed to diffuse 
with a coefficient of diffusion $D$. 
The last dynamical rule defining the model consists of deposition events.
When particles have diffused in such a way that a void of unit size opens, 
the void is instantaneously filled by a new particle
and the density, $\rho(\tw)=N(\tw)/L$, increases by $1/L$, $N(\tw)$
being the total number of particles present at time $\tw$. 
These rules lead to slow dynamics, because 
the larger the density the longer it takes to open a void.
A more general version of this model
includes evaporation~\cite{kra,dep}. It 
displays the generic features observed in granular compaction:
logarithmic increase of the density~\cite{kra,sti,dep},
aging~\cite{dep,tal},
non-Gaussian density fluctuations~\cite{kol},
effective temperatures and link with Edwards entropy~\cite{dep}, 
hysteresis effects~\cite{kni}. 
The time evolution of the density also 
describes well the experiments,
\begin{equation}
\rho(\tw)=
\rho_{\infty}- \left[ a_0+a_1\ln\left(\frac{\tw-t_0}{\tau}\right) 
\right]^{-1},
\label{eqn:comp} 
\end{equation}
where $\rho_{\infty}$, $a_0$, $a_1$, $\tau$ and $t_0$ are fitting
parameters. For the present model without evaporation 
Eq.~(\ref{eqn:comp}) holds at large $\tw$ with 
$\rho_{\infty}= a_1=\tau=1$, $a_0=t_0=0$. Let us remark here that such 
logarithmic compaction has also been obtain from different models 
with deposition,  
evaporation and diffusion, like those studied in Ref.~\cite{majd}.
Here we focus on local dynamic 
quantities and their spatial correlations through both analytic calculations
and Monte Carlo simulations. The results of the simulations have been
obtained by averaging over $2.10^4$ independent histories with 
$L=250$. Time is measured in units of the diffusion constant, $D$,
which is set to unity.

\begin{figure}
\begin{center}
\psfig{file=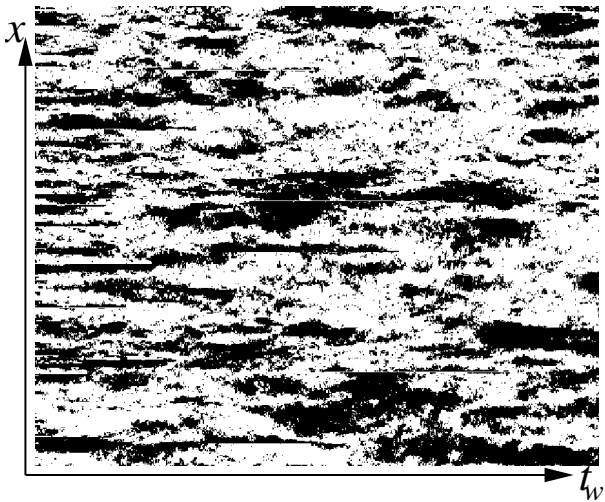,width=8.cm}
\end{center}
\caption{\label{field} Space-time
pattern of dynamic fluctuations: particles are represented
in black (white) when moving slower (faster) than average. 
Local dynamics is measured via the self-intermediate scattering function, 
Eq.~(\ref{fs}), with $k=\pi$, at fixed $t-\tw=2.10^4$.}
\end{figure}

In Fig.~\ref{field} we illustrate the results 
reported in this paper, namely the emergence of spatio-temporal 
correlations in dynamic trajectories of 
compacting granular systems. In this figure, 
black (white) denote particles that 
are more (less) mobile than the average
on a timescale of the order of the structural relaxation time.  
The time extension of the trajectory is $10^5$, 
the spatial extension $L=500$, and the final density 
$\rho(10^5) \approx 0.91$.
The emergence of large-scale correlations 
in the local dynamics is evident, since 
domains that extend spatially over a hundred particles
can be observed, despite the absence of any such large
static correlations between particles.   
Spatial clustering of fast and slow regions 
is the hallmark of dynamic heterogeneity~\cite{dh,wee,kcm,mayer}. 
A closer look at the left part of the figure shows that the 
dynamic lengthscale is visibly smaller at shorter times, 
suggesting that spatial correlations are larger when 
the dynamics becomes slower, and possibly diverge 
in the jamming limit  of a full system.

How did we build Fig.~\ref{field}? 
As in supercooled liquids, information on local dynamics 
is accessed by following the dynamics of individual particles. 
Consider two times $\tw$ and $t>\tw \gg 1$.
The distribution of particle displacements 
is the self-part of the van-Hove function, 
\begin{equation}
G_s(r,t,\tw)=\frac{1}{N(\tw)} \sum_{i=1}^{N(\tw)} 
\left\langle \delta \big( r-\delta r_i(t,\tw) \big) \right\rangle,
\label{vanhove}
\end{equation}
where $\delta r_i(t,\tw)=r_i(t)-r_i(\tw)$ 
is the displacement of particle $i$ between 
times $\tw$ and $t$ and
only particles present at time 
$\tw$ are summed over.
Its Fourier transform is the self-intermediate scattering function  
\begin{equation}
F_s(k,t,\tw)=\frac{1}{N(\tw)}\sum_{i=1}^{N(\tw)} 
\left\langle \cos \big( k \, \delta
r_i(t,\tw) \big) \right\rangle.
\label{fs}
\end{equation}
Aging will manifest itself through
an explicit dependence of $F_s(k,t,\tw)$ on its two time
arguments~\cite{reviewaging}. In the following we 
focus on the value $k = 2 \pi$ which corresponds 
to displacements of the order of the particle size.

Since dynamic heterogeneity relates 
to spatial fluctuations about the averaged 
two-time dynamics, we have shown in Fig.~\ref{field} 
spatial fluctuations about $F_s(k,t,\tw)$.
We colored black (white) those particles for which
$\delta F_i(k,t,\tw) = 
\cos \big( k\delta r_i(t,\tw) \big) - 
F_s(k,t,\tw)$ is negative (positive), 
i.e. those particles that move more (less) than average
in a particular realization.

To further quantify how particle displacements are spatially correlated
we consider the structure factor
of the dynamic heterogeneity shown in Fig.~\ref{field}, 
\be
S_k(q,t,\tw) = \frac{1}{N(\tw)} 
\sum_{l,m=1}^{N(\tw)}
\frac{ \delta F_l
\delta F_m}{f_k(t,\tw)}
e^{i q [r_l(\tw) - r_m(\tw)]} \, ,
\label{s4}
\ee
where
$f_k(t,\tw) = \langle \delta F_i(k,t,\tw)^2 \rangle$ normalizes
the structure factor and some obvious time and 
wavevector dependencies have been 
removed, for clarity. By definition, the structure factor 
is built from two-time two-point quantities
which is the minimum requirement to 
probe the spatio-temporal patterns of Fig.~\ref{field}.
The small $q$ limit in Eq.~(\ref{s4})
defines a dynamic ``four-point'' susceptibility, 
\be
\chi_k(t,\tw) = S_k(q=0, t,\tw), 
\label{chi4}
\ee
which can be rewritten as the variance of the fluctuations
of the spatially averaged two-time dynamics. 
Physically, dynamic fluctuations increase when the number of
independently relaxing objects decreases~\cite{mayer}. Normalizations
ensure that $\chi_k$ is finite in
the thermodynamic limit, except at a dynamic critical 
point~\cite{steve}.

The quantities introduced above can be estimated analytically
as follows. 
Between two deposition events, the total free volume is $L(1-\rho)$
and thus the slow dynamics can be shown~\cite{stile} to be equivalent
to a continuum version of a 
symmetric exclusion process at density $\rhot=\rho/(1-\rho)$,
i.e. involving caged dynamics.
This process is then
mapped to a fluctuating interface model~\cite{stile,maj,alex}. 
Labeling particles in the order of their position 
from the origin
we introduce an interface position  $y(x,t)$ for
the $i$th particle encountered on the lattice, such that
$y(x,t)=r_i(t)$, $x$ replacing $i$ as a label for the
particle. 
We then write $y(x,t)=h(x,t)+x/\rho$, were $h(x,t)$
is the deviation from the position of the particle from an 
uniform configuration. The connection to 
symmetric exclusion process allows one to show~\cite{stile,maj,alex} 
that $h(x,t)$ follows an Edwards-Wilkinson dynamics~\cite{edw}, 
\begin{equation}
\frac{\partial h}{\partial t}
 = \alpha\nabla^2 h+\eta,
\label{eqn:EW}
\end{equation}
where $\alpha=\rhot^2\simeq\ln(\tw)^2$ and
$\eta$ is a Gaussian white noise with
zero mean and variance
$\langle \eta(x,t)\eta(x',t')\rangle=2\delta(x-x')\delta(t-t')$.
From (\ref{eqn:EW})
the equilibrium measure reads
\begin{equation}
P_{\rm eq}[h(x)]=\exp \left( - \frac{1}{2\alpha} \int_{-\infty}^\infty 
dx \, \left[\nabla h(x)\right]^2 \right),
\label{eqn:p}
\end{equation} 
which shows that 
deviations from the averaged position of a particle 
arise with an elastic penalty in the interface representation.
This elastic behavior makes the following 
calculation similar to the evaluation 
of the elastic contribution to the
dynamical susceptibility (\ref{chi4}) 
in supercooled liquids performed in Ref.~\cite{bir}.

Equation~(\ref{eqn:EW}) is first solved in the Fourier space, 
\begin{equation}
\hat{h}(q,t)=e^{-\alpha q^2 \tau}\ \hat{h}(q,\tw)+\int_{\tw}^t dt' \, 
\eta(q,t') e^{-\alpha(t-t')},
\label{eqn:h}
\end{equation} 
where we have defined the time difference $\tau = t-\tw$.
A crucial approximation is made here
since $\alpha$ in Eq.~(\ref{eqn:h}) has in fact 
a logarithmic time dependence. 
This amounts to neglecting 
the effect of further deposition events on 
the particles already present at $\tw$.
Since deposition is such a rare event 
at large times this approximation should capture the 
evolution of the local dynamics, as our
numerical simulations shall confirm.
Using (\ref{eqn:h}), one easily gets 
\begin{equation}
F_s(k,t,\tw)=\exp \left
( -k^2\sqrt{\frac{2 \tau}{\alpha(\tw)}}g(0)\right),\label{eqn:fs}
\end{equation}
where
$g(x)=\int_{-\infty}^\infty  dq \,
e^{iqx} (1-e^{-q^2/2})/ q^2$, 
so that $g(0)=(2\pi)^{1/2}$. 
Equation (\ref{eqn:fs}) is a
classic result for the ordering process 
of one-dimensional random walkers~\cite{har,arr}.
It shows that $F_s(k,t,\tw)$ displays aging behavior 
and scales with $\tau / \tau_k(\tw)$, 
with a logarithmically increasing
relaxation timescale $\tau_k(\tw) \sim \alpha(\tw) / k^4$.
Particles are therefore sub-diffusing, 
$\langle \delta
r_i^2(t,\tw)\rangle \sim  \sqrt{\tau/\alpha(\tw)}$, and
the van Hove function (\ref{vanhove}) is a Gaussian.
Gaussiannity is a consequence of the diffusive 
nature of the particle motion at short times, which
explains the discrepancy 
with the non-Gaussian distributions of displacements 
that have recently been measured experimentally 
in a bidimensional geometry~\cite{mar}.

The dynamic susceptibility (\ref{chi4}) 
can also be computed,
\begin{eqnarray}
\chi_k(t,\tw) & = & \int dx\
\frac{\cosh\left(\frac{2 k^2}{\rhot(\tw)}\sqrt{2 \tau}
  g(\frac{x}{\rhot(\tw)\sqrt{2\tau}})\right)-1}{ \cosh 
\left(\frac{2 k^2}{\rhot(\tw)} \sqrt{2
\tau} g(0)\right)- 1 } \nonumber \\
& = &\frac{\alpha(\tw)}{k^2} {\cal F}\left(\frac{\tau}{\tau_k(\tw)}\right),
\label{eqn:chi}
\end{eqnarray}
where ${\cal F}(x)$ is a scaling function 
defined from the first line of Eq.~(\ref{eqn:chi}).  
Careful analysis of Eq.~(\ref{eqn:chi}) shows that
$\chi_k(t,\tw)$ goes from zero at $\tau=0$ to 
the asymptotic value $\chi_k(t\to\infty,\tw) = \alpha(\tw) / 
(2 k^2 |g'(0)| )$, via a maximum
$\chi_k^\star \sim \alpha(\tw)/ k^2$ when 
$\tau \sim \tau_k(\tw)$. 

\begin{figure}
\begin{center}
\includegraphics[width=.45\textwidth]{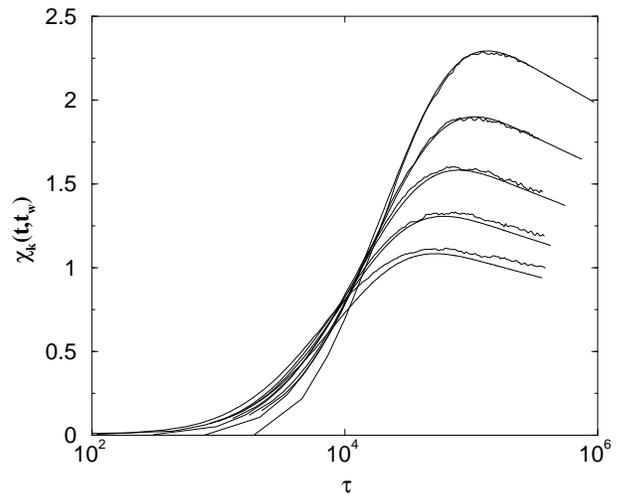}
\caption{Comparison of the numerical and analytical evaluations of 
$\chi_k(t,\tw)$ as a function of $\tau=t-\tw$
for $k = 2 \pi$ and  
$\tw=6250$, $12500$, $25000$, $50000$, $100000$ (from bottom to top).}
\label{fig:chi4}
\end{center}
\end{figure}

In Fig.~\ref{fig:chi4} we show the dynamic 
susceptibility as a function of time separation $\tau$
for various ages $\tw$ obtained from direct 
numerical simulations. The lines through the data are
from our analytical estimation, Eq.(\ref{eqn:chi}), 
which constitutes clearly an excellent 
quantitative representation of the data
when $\tw$ becomes large. This supports 
the approximation made above that the dynamics 
at large times is slow enough that deposition events 
have little influence on local dynamics. 

The curves in Fig.~\ref{fig:chi4} 
strikingly resemble the four-point susceptibilities 
discussed for supercooled liquids approaching their glass 
transition~\cite{dh}
and simple coarsening systems~\cite{mayer}.
As for those systems, we conclude that dynamics is 
maximally heterogeneous when observed 
at time separations close to the relaxation timescale, 
itself dependent of the age of the sample.
To our knowledge no experimental determination of 
the dynamic susceptibility $\chi_k(t,\tw)$ 
has been reported for granular media, 
although the experimental set-up 
described in Ref.~\cite{mar} would probably allow its determination.

\begin{figure}
\begin{center}
\includegraphics[width=.45\textwidth,clip]{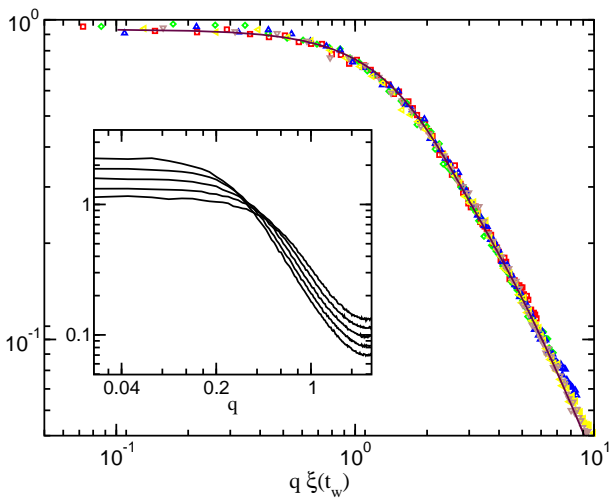}
\caption{Inset: Simulated dynamic structure factor (\ref{s4}) as a function of
$q$ for parameters as in Fig.~\ref{fig:chi4} and for $\tau=\tau_k(\tw)$.
Main: Dynamic scale invariance, Eq.~(\ref{scale_invariance}), 
is revealed by rescaling space by $\xi_k$ and $S_k$ by
$\chi_k^*$.  The full line is the analytical
result using the mapping to the Edwards-Wilkinson interface model.} 
\label{fig:s4}
\end{center}
\end{figure}

The dynamics susceptibility $\chi_k(t,\tw)$ 
measures the volume integral of a spatial 
correlator. Therefore, an increasing susceptibility 
directly suggests the existence of a growing dynamic correlation length. 
This is most directly seen in the Fourier space when the 
wavevector dependence of the dynamic structure factor 
(\ref{s4}) is considered. 
We have obtained an analytical form 
for $S_k(q,t,\tw)$ but it is too lengthy to be reported 
here~\cite{stile}. In 
Fig.~\ref{fig:s4} we show the wavector dependence of 
$S_k(q,t,\tw)$ for various $\tw$ at time separations corresponding 
to the maximum of the dynamic susceptibility, obtained in numerical
simulations. We also show the analytical results.
As for the dynamic susceptibility the agreement between 
analytical and numerical results is excellent.  
   
At fixed $\tw$ the structure factor is characterized by a plateau 
at small $q$ whose height is given by $\chi_k^\star(\tw)$.
When $q$ increases $S_k(q,t,\tw)$ leaves the plateau and decreases
to 0 at large $q$.
When $\tw$ is increased the plateau becomes higher and
it ends at a smaller wavevector but 
its overall shape is unchanged. This implies 
a dynamic scale invariance such as found in 
glass-formers~\cite{kcm,steve}:
rescaling times by $\tau_k(\tw)$
and space by $\xi_k(\tw) \sim \alpha(\tw) / k^2$ makes trajectories
statistically equivalent. Formally this means that
the following scaling law is obeyed, 
\be
S_k(q, \tau_k, \tw) = \chi_k^\star(\tw) \, 
{\cal G}_k \left( q \xi_k(\tw) \right).
\label{scale_invariance}
\ee 
Again we note that data in Fig.~\ref{fig:s4} strikingly resemble 
dynamic structure factors measured both in realistic 
supercooled liquids~\cite{dh} 
and coarse-grained models for the glass transition~\cite{kcm}.

A major result of the above analysis is the existence
of a dynamic length scale, $\xi_k(\tw)$, which
grows logarithmically with time when compaction proceeds, and therefore
diverges when the systems jams.
A diverging lengthscale provides 
support to the temporal renormalization group argument 
developed in Ref.~\cite{sti}, but we see no obvious
connection between the dynamic criticality described here and the various
power law scalings observed in static systems approaching jamming from
above~\cite{ohern}.
Physically, collective rearrangements of
particles are needed to create a void of unit size, the
denser the system the more cooperative the dynamics.
A naive determination of $\xi$ would rely on 
a free volume argument~\cite{bou}. 
The mean free volume available to particle is $1/\rhot$, so that
the number of particles required to have a fluctuation 
of the free volume equal to unity is ${\cal N} \sim \rhot^2 \sim 
\xi(\tw)$. This simple physical argument underlies the cooperative 
nature of the dynamics which is more formally 
captured by four-point correlation 
functions, Eqs.~(\ref{s4}) and (\ref{chi4}).

The idea that the size of collective motions 
increases when dynamics becomes slow is certainly not new. 
Multi-point correlation functions have now been 
measured in very different materials with similar qualitative results. 
That large dynamic lengthscales control glassy dynamics
suggests the possibility that few universality classes 
underly and possibly unify the dynamical behavior of a 
much wider diversity of jamming materials. 

\begin{acknowledgments}
We thank S.N. Majumdar for enlightening discussions. 
\end{acknowledgments}

\end{document}